\documentclass{article}

\usepackage{arxiv}

\usepackage{amsmath, amssymb} 
\usepackage{hyperref} 
\usepackage{array} 
\usepackage{multirow}
\usepackage{caption}
\usepackage{subcaption}
\usepackage{cleveref}

\usepackage{graphicx}
\graphicspath{{/Users/williazo/Documents/UCLA/DissertationDefense/OUTPUT/IMAGES/}{./}}

\usepackage[square, comma, numbers, sort&compress, sectionbib]{natbib} 

\usepackage[flushleft]{threeparttable} 
\usepackage{xcolor, colortbl} 
\definecolor{dodgerblue4}{RGB}{16, 78, 139}
\definecolor{dodgerblue1}{RGB}{30, 144, 255}
\definecolor{indianred4}{RGB}{139, 58, 58}
\definecolor{indianred1}{RGB}{255, 106, 106}
\definecolor{viridis1}{RGB}{68, 1, 84}
\definecolor{viridis1d}{RGB}{68, 1, 84}

\DeclareMathOperator*{\E}{\mathbb{E}}
\DeclareMathOperator*{\C}{\mathbf{\textit{C}}}

\DeclareMathOperator{\indp}{\perp \!\!\! \perp}

\title{Causal inference for multiple continuous exposures via the multivariate generalized propensity score}

\author{
	Justin R.~Williams\\
	Department of Biostatistics\\
	University of California, Los Angeles\\
	Los Angeles, CA, USA, 90049
\And
	Catherine M.~Crespi\\
	Department of Biostatistics\\
	University of California, Los Angeles\\
	Los Angeles, CA, USA, 90049}

\renewcommand{\headeright}{}

\begin{document}
\maketitle

\begin{abstract}
The generalized propensity score (GPS) is an extension of the propensity score for use with quantitative or continuous exposures (e.g., dose of medication or years of education). Current GPS methods allow estimation of the dose-response relationship between a single continuous exposure and an outcome. However, in many real-world settings, there are multiple exposures occurring simultaneously that could be causally related to the outcome. We propose a multivariate GPS method (mvGPS) that allows estimation of a dose-response surface that relates the joint distribution of multiple continuous exposure variables to an outcome. The method involves generating weights under a multivariate normality assumption on the exposure variables.  Focusing on scenarios with two exposure variables, we show via simulation that the mvGPS method can achieve balance across sets of confounders that may differ for different exposure variables and reduces bias of the treatment effect estimates under a variety of data generating scenarios. We apply the mvGPS method to an analysis of the joint effect of two types of intervention strategies to reduce childhood obesity rates.
\end{abstract}


\section{Introduction}\label{intro}
Analyzing data from observational studies and non-randomized experiments to estimate causal effects is challenging due to potential confounding between the exposure and outcome. Techniques such as propensity score methods can help to remove sources of potential confounding and return valid estimates of the treatment effect \citep{rosenbaum_propensity}. Propensity score methods were first developed for binary treatments, and then extended to categorical, or multiple, treatments, which introduced the term ``generalized propensity score" (GPS) \citep{imbens_dose_resp}. In the context of a categorical treatment variable, the GPS corresponds to the conditional probability of receiving a particular treatment given a set of confounders. Following the extension to categorical treatments, the GPS was adapted to the setting of continuous exposures via the use of conditional densities \citep{hirano_continuous, imai_causalGPS}. 

Originally, the GPS for continuous exposures was estimated using Gaussian densities, with adjustment for confounding accomplished through either covariate regression \citep{hirano_continuous} or stratification \citep{imai_causalGPS}. Several recent methods have aimed at increasing flexibility for estimating the GPS by using gradient boosting \citep{zhu_boosting}, kernel smoothing \citep{flores2012, kennedy2017}, or ensemble algorithms \citep{kreif_gps_ml}. Other methods focus on the use of weights \citep{robins2000marginal}, including incorporating covariate balancing properties in estimation using a penalized likelihood approach \citep{fong2018} or constrained optimization on the entropy of the weights \citep{tbbicke2020entropy, vegetabile2020nonparametric}. All of these methods have maintained the assumption that the exposure is univariate, i.e., a single continuous treatment variable. Methods to accommodate multiple simultaneous treatment exposures have been only briefly mentioned in the literature \citep{imai_causalGPS}. 

There are many situations in which evaluating the combined effect of multiple simultaneous exposures is critical to answering scientific questions. In medicine, combination therapies, which involve the patient taking several medications simultaneously, have been shown to be effective for treating Crohn's disease \citep{crohns_combo_tx}, cancer \citep{cancer_combo_tx}, hypertension \citep{hypertension_combo_tx}, and HIV \citep{hiv_combo_tx}. Typical methods for estimation of the dose response surface for combination treatments requires careful design with repeated randomized experiments in order to estimate the optimal combination of treatment doses \citep{khuri_rsm}. When such experimentation is not feasible, researchers may wish to use available data from observational or non-randomized studies to estimate the joint effects. For example, there is currently interest in studying potential combination therapies for COVID-19 using available data from non-randomized studies. However, determining a potentially beneficial dose of several medications may be complicated due to confounding by patient demographic characteristics, comorbities or other factors \citep{covid_combo_stebbing, covid_combo_gautret, covid_combo_gautret2, covid_tx_review}.

In this article we develop methods to estimate the causal effects of multiple continuous exposures occurring simultaneously, using data from a non-randomized study. We develop a general framework for estimating the causal effects of a multivariate exposure of arbitrary dimension, but focus on bivariate exposures in our simulations and motivating example. The primary objective is unbiased estimation of the dose-response surface of the average outcome given a particular combination of exposure values. We propose methods for estimating weights using a multivariate generalized propensity score, which we call mvGPS, and use weighted regression to estimate the dose-response surface. Our methods rely on the assumption that the exposure variables have a multivariate normal distribution.

In Section~\ref{background}, we introduce our motivating example, which involves assessing the joint effect of two types of intervention strategies for reducing childhood obesity rates in Los Angeles County. In Section~\ref{methods}, we propose a method of causal inference with multiple simultaneous continuous exposures using the multivariate generalized propensity score. Section~\ref{sim_section} presents a simulation study designed to highlight strengths and limitations of the methodology. Section~\ref{wic_results} applies the proposed methods to our motivating example. A discussion in Section~\ref{discussion} concludes the paper. 

\section{Motivation}\label{background}
Obesity rates among low-income preschool-aged children in Los Angeles County were consistently higher than the national average in 2003-2009, with about 20\% of such children classified as obese (BMI $\ge$ 30 kg/m$^2$) \citep{phfewic}. In response, several organizations, including Los Angeles County Department of Public Health, First 5 LA, Nemours, and the Special Supplemental Nutrition Program for Women, Infants and Children (WIC), implemented programs and policies aimed at reducing childhood obesity in the county. The interventions used a wide variety of different approaches and reflected a large investment of resources.

The Early Childhood Obesity Systems Science (ECOSyS) study, funded by National Institute of Health R01 HD072296, sought to evaluate the impact of these programs on childhood obesity prevalence. To this end, ECOSyS collected information on the nature, timing, location and reach of programs implemented in the county in 2003-2016. The research team also developed a method of calculating a ``community intervention dose index” that aggregates exposure to childhood obesity interventions over multiple different programs \citep{wang2018developing}. The community intervention dose index is calculated using a multistep procedure. Each program is coded to location and year of implementation, extent of reach into the target population, and which of nine different intervention strategies it used. The nine strategies are listed in Table~\ref{tab:strat_tbl}. The strategies are categorized as ``micro” strategies, which target specific individuals, or ``macro” strategies, which target a population at large. By aggregating over the strategies implemented in a particular location during a particular year, strategy-specific as well as total micro and macro intervention dose indices can be calculated. 

For this paper, we focus on intervention exposures stemming from WIC programs. WIC serves low-income families and has seven agencies within Los Angeles County with approximately 90 clinics. While WIC offers many regular services, primarily food assistance and nutrition education that are uniform from clinic to clinic, WIC agencies can also receive additional funding to implement intervention programs. These programs are implemented non-randomly at clinics due to differences in community needs and other considerations. Our motivating example focuses on WIC intervention programs implemented in 2010-2016, given that a major change in the WIC food package occurred in 2009 that may have altered family behaviors and neighborhood food environments \citep{food_pacakage_change, hillier2012impact}.

A total of 32 WIC intervention programs implemented in Los Angeles County from 2010-2016 were cataloged by the ECOSyS research team. Information about each program was obtained, including how it was implemented, the estimated reach in terms of client participation, which clinics participated, and how long the program was active. These exposures were mapped to census tracts where WIC participating children live by identifying the implementing WIC clinics and then defining a catchment area around the clinic intended to capture the exposed population. Catchment areas were defined for each clinic using records of client attendance, and varied by strategy type (macro vs. micro). Children living in census tracts that fell within a catchment area were assumed to be exposed. Exposure values at census tracts were then aggregated across programs by strategy and by year to obtain a single continuous dose for each of the nine intervention strategies. Strategy-specific doses were then summed into macro and micro intervention doses, which were log transformed due to skewness. More details about the dose construction procedure are provided in a forthcoming paper. 

\begin{figure}[ht]
\centering
\caption{Joint Distribution of Macro and Micro Intervention Doses}
\includegraphics[width = \textwidth, keepaspectratio]{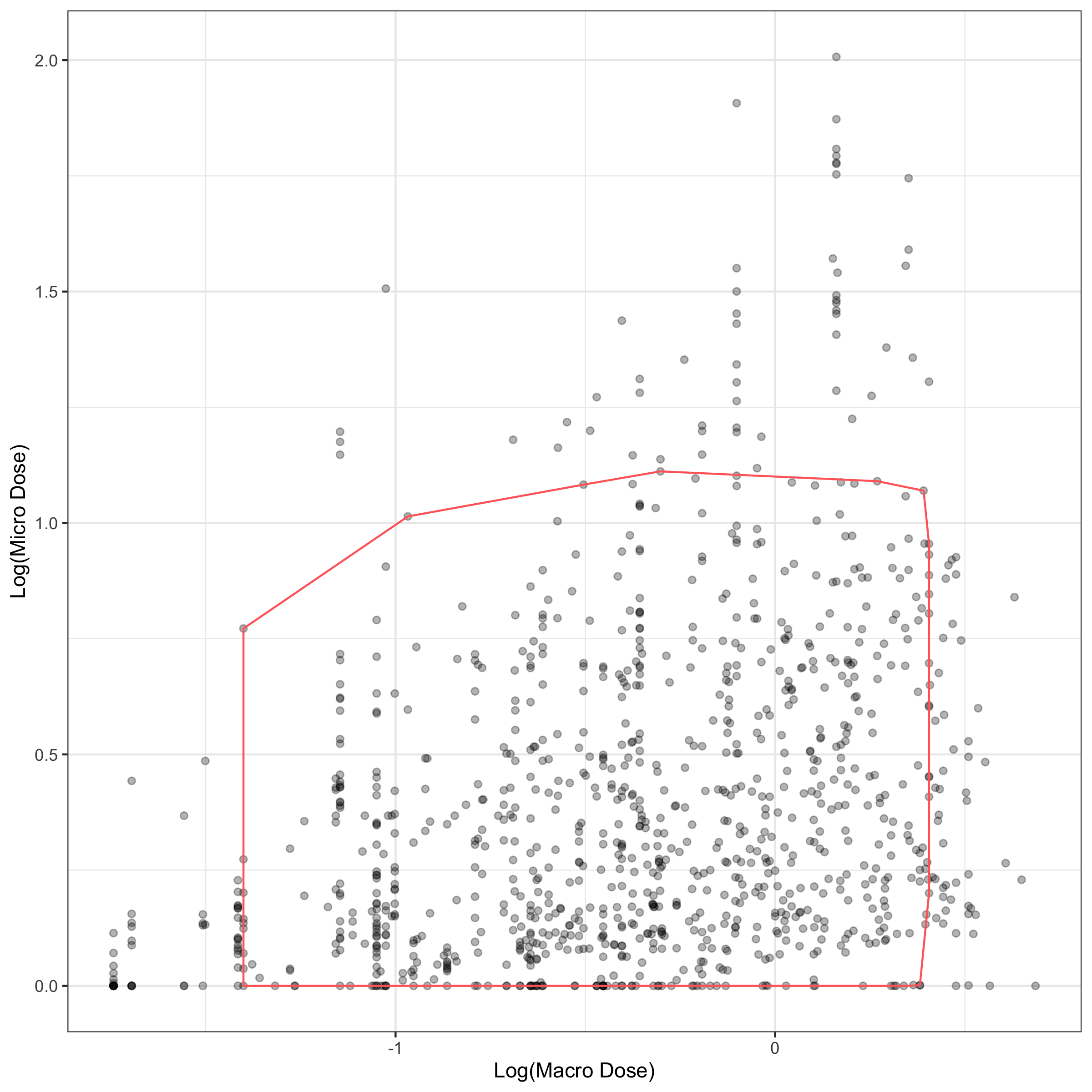}
\caption*{Observed values of $\log$ macro and micro exposure doses averaged over the study intervention period 2010-2016. Trimmed convex hull region with $q=0.95$, i.e., $\mathcal{H}_{0.95}$, is shown in \textcolor{indianred1}{red}}
\label{fig:macro_micro_joint}
\end{figure}

Figure~\ref{fig:macro_micro_joint} shows the resultant joint distribution of macro and micro intervention dose for the WIC intervention programs averaged over our defined intervention period, 2010-2016. Each point in the figure represents a census tract. WIC-participating children residing in a particular census tract were presumed to receive the calculated doses.

The outcome of interest was change in census tract-level childhood obesity prevalence. Childhood obesity prevalence was measured using administrative records from children ages 2-5 years who participated in WIC in Los Angeles County during 2007-2016, compiled by the WIC Data Mining Project, see \url{https://lawicdata.org} for more details. From these records, obesity prevalence by census tract and year was constructed for census tracts with at least 30 WIC-enrolled children. Census tracts used in the analysis were restricted to 8 regions within Los Angeles County that were targeted as part of the ECOSyS data collection effort. This resulted in a total of $n=1079$ census tracts which serve as the units of analysis. The outcome of interest, $Y$, was the difference in average obesity prevalence between post, 2012-2016, and pre, 2007-2009, intervention, i.e., $Y=\bar{p}_{post}-\bar{p}_{pre}$. The post intervention period was taken to start in 2012 rather than 2010 to account for potential lag in treatment effects.

We aimed to estimate the dose-response surface of Y, change in childhood obesity prevalence, associated with combinations of macro and micro intervention exposure doses, after removing bias due to non-random assignment of programs. Understanding the simultaneous effect of macro and micro intervention strategy exposures is important to help policy makers make decisions about the allocation of scarce resources to various intervention strategies \citep{rosenkranz2008}.

\section{Methods}\label{methods}
\subsection{Notation}\label{subsec:methods_notation}
Our approach to developing the multivariate generalized propensity score follows the Neyman-Rubin causal model and uses the potential outcome notation introduced by Neyman \citep{neyman1923} and made popular by Rubin \citep{rubin1974}. Let $Y_i$ denote the outcome of interest for unit $i$ from a population of size $n$ and $\mathbf{D}_i$ be a vector of length $m$ providing the values for $m$ continuous exposures for unit $i$. The confounders relevant to each exposure are allowed to be different. Let $\C_i=\{\mathbf{C}_{i1},\dots, \mathbf{C}_{im}\}$ be a set of size $m$ where each element in the set, $\mathbf{C}_{ij}$, $j=1,\dots,m$, is a $p_{j}$ dimensional vector of baseline confounders associated with the $j^{th}$ exposure and the outcome. We denote the value of the $k^{th}$ confounder of the $j^{th}$ exposure for the $i^{th}$ individual as $C_{ijk}$, with $i=1,\dots, n$, $j=1,\dots, m$, and $k=1,\dots, p_{j}$. If all exposures have identical confounders, then $\mathbf{C}_{i1}=\cdots=\mathbf{C}_{im}=\mathbf{C}_{i}$ and $p_{1}=\cdots=p_{m}=p$. The observed data for the $i^{th}$ unit is represented as $(Y_{i}, D_{i1},\dots,D_{im}, C_{i11}, \dots, C_{i1p_{1}}, \dots, C_{im1}, \dots, C_{imp_{m}})$. Further, we define the potential outcome $Y_{i}(\mathbf{d})$ as the outcome that the $i^{th}$ subject would have if assigned the exposure vector $\mathbf{d}=(d_{1},\dots,d_{m})$. We will use capital $\mathbf{D}$ to represent the multivariate random variable representing dose combinations, and lowercase $\mathbf{d}$ as a particular value in the multidimensional space. Estimation focuses on the average dose-response function defined as $\mu(\mathbf{d})=\E[Y(\mathbf{d})]$, which is assumed to be well defined for any $\mathbf{d}\in\mathcal{D}\subseteq\mathbb{R}^{m}$. Note that with a bivariate exposure, i.e., $m=2$, $\mu(\mathbf{d})$ is a dose-response surface in 3-dimensional space.

\subsection{Identification Assumptions}\label{subsec:id_assump}
We make the following identifying assumptions: weak ignorability, positivity, and stable-unit treatment value. Weak ignorability, also known as selection on observables or unconfoundedness, states that exposure is conditionally independent of the potential outcomes given the appropriate set of confounders. We write this in the multivariate case as
\begin{equation*}
Y_{i}(\mathbf{d})\indp\mathbf{D}_{i}\mid \mathbf{C}_{i1}, \dots, \mathbf{C}_{im} \quad \forall \quad \mathbf{d}\in\mathcal{D}.
\end{equation*}
When this assumption holds, we can replace the high-dimensional conditioning set with a scalar value by means of the conditional density function of exposure \citep{rosenbaum_propensity}. In our case the conditional density is defined as the multivariate generalized propensity score (mvGPS), which we denote $f_{\mathbf{D}\mid \mathbf{C}_{1},\dots,\mathbf{C}_{m}}$. Weak ignorability is often the most difficult assumption to rationalize as it requires perfect knowledge and collection of all possible confounders of the exposures and outcome in the set $\C$. We assume that the set $\C$ is well defined and that there is no unmeasured confounding.

\begin{figure}[ht]
\centering
\caption{Defining Estimable Region with Bivariate Exposure}
\includegraphics[width = \textwidth, keepaspectratio]{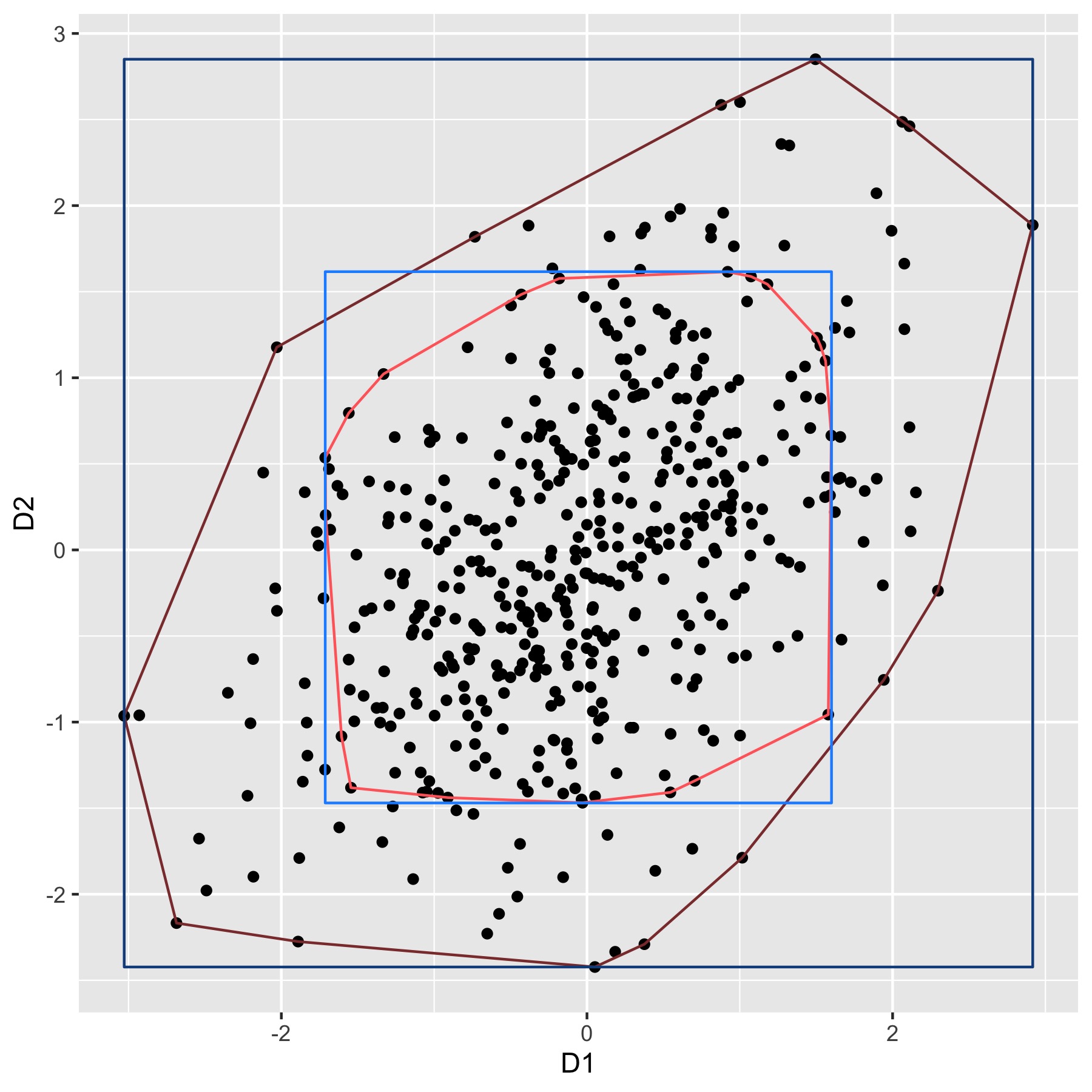}
\caption*{Sample of $n=500$ units drawn from population where $D_{1}\sim\mathnormal{N}(0, 1), D_{2}\sim\mathnormal{N}(0, 1), Cov(D_{1}, D_{2})=0.5$. Region defined by \textcolor{dodgerblue4}{dark blue} box corresponds to $\mathcal{G}$ while region defined in \textcolor{indianred4}{dark red} represents $\mathcal{H}$. Trimmed regions are also shown for $q=0.95$ with the \textcolor{dodgerblue1}{light blue} box corresponds to $\mathcal{G}_{0.95}$ while the region defined in \textcolor{indianred1}{light red} represents $\mathcal{H}_{0.95}$}
\label{fig:chull}
\end{figure}

The second assumption, positivity, claims that all units have the potential to receive a particular level of exposure given any value of the confounders. In notation, we have 
\begin{equation*}
0<f_{\mathbf{D}\mid \mathbf{C}_{1},\dots,\mathbf{C}_{m}}(\mathbf{D}=\mathbf{d}\mid \mathbf{C}_{1},\dots,\mathbf{C}_{m})<1  \quad \forall \quad \mathbf{d}\in\mathcal{D}.
\end{equation*}
This assumption requires that we carefully define $\mathcal{D}$ such that all units have the potential to receive any particular value in the domain. In the case of a univariate continuous exposure, positivity is often enforced by restricting estimation to either the observed range or a trimmed version \citep{crump2009dealing}. For example, using the observed range we would define $\mathcal{D}=[d_{0}, d_{1}]$ where $d_{0}$ and $d_{1}$ correspond to the minimum and maximum observed exposure. In the case of a multivariate exposure, a natural inclination might be to extend this approach to multiple dimensions by setting $\mathcal{D}=\mathcal{G}$ where $\mathcal{G}$ is defined as
\begin{equation*}
\mathcal{G}=\prod_{j=1}^{m}[d_{0j}, d_{1j}]\subset\mathbb{R}^{m},
\end{equation*}
where $d_{0j}$ and $d_{1j}$ are the minimum and maximum observed exposure, respectively, along dimension $j$. However, when exposure variables are correlated, i.e., $Cov(D_{j}, D_{j'})\ne0$ for $j\ne j'$, the region $\mathcal{G}$ may include areas with few or no observations. Instead, we propose defining the estimable region for multivariate exposures as $\mathcal{D}=\mathcal{H}\subset\mathcal{G}$, where $\mathcal{H}$ is defined as the convex hull of the multivariate exposure \citep{chazelle1993optimal}. Using a convex hull ensures that inference is restricted to regions where data are observed and avoids extrapolating to sparse data regions in the multidimensional space. For the case of $m=2$, Figure~\ref{fig:chull} shows the difference between regions $\mathcal{G}$ and $\mathcal{H}$ when $Cov(D_{1}, D_{2})=0.5$. Additionally, similar to the univariate case, we can define trimmed versions of $\mathcal{G}$ or $\mathcal{H}$. By specifying a value $q\in[0.5, 1]$, we construct $\mathcal{G}_{q}$ using trimmed minimum and maximum values as
\begin{equation*}
\mathcal{G}_{q}=\prod_{j=1}^{m}[d^{q}_{0j}, d^{q}_{1j}]\subset\mathcal{G},
\end{equation*}
where $d_{0j}^{q}=Q(\mathbf{d}_{j}, 1-q)$, $d_{1j}^{q}=Q(\mathbf{d}_{j}, q)$, and $Q(\cdot, q)$ is the sample quantile function. To create the trimmed convex hull, $\mathcal{H}_{q}$, we recalculate the convex hull using the subset of observations that falls within the trimmed minimum and maximum across all exposure dimensions.

The final assumption is the stable-unit treatment value assumption (SUTVA), which states that the potential outcome of each unit does not depend on the exposure that other units receive and that there exists only one version of each exposure \citep{rubin1980}. This assumption rules out potential interference between units or other errors in defining the potential outcomes caused by multiple versions of the exposure. Therefore the potential outcomes are well-defined for each unit and the observed outcome given exposure $\mathbf{D}=\mathbf{d}$ corresponds to the potential outcome, i.e., $Y_{i}(\mathbf{d})=Y_{i}$. We discuss the tenability of this assumption to our data application in the Discussion.

\subsection{Multivariate Generalized Propensity Score}\label{subsec:mvGPS}
Using the identifying assumptions above, there are a variety of different methods to estimate the dose-response function including covariate adjustment \citep{hirano_continuous} or stratification \citep{imai_causalGPS}. We focus on weighted estimation, originally proposed for binary treatments with marginal structural models \citep{robins2000} and motivated by weights used in survey sampling \citep{horvitz1952}. We aim to construct a set of weights, $w$, that when applied to the observed data return a consistent estimate for the average dose-response function, i.e.,
\begin{equation}\label{eqn:unbias_w}
    \E[wY\mid\mathbf{D}]=\E[Y(\mathbf{d})].
\end{equation}
In the case of univariate continuous exposure, weights are constructed by either estimating the generalized propensity score \citep{hirano_continuous, imai_causalGPS, kennedy2017, zhu_boosting, fong2018} or by direct optimization using an entropy loss function \citep{tbbicke2020entropy, vegetabile2020nonparametric}. We choose to extend the generalized propensity score by using an appropriately defined multivariate conditional distribution, which we refer to as the multivariate generalized propensity score (mvGPS). The weights are thus constructed as the ratio of the multivariate marginal density to the conditional density
\begin{equation}\label{eqn:mvGPS_wts}
w=\frac{f(\mathbf{D})}{f(\mathbf{D}|\mathbf{C}_{1},\dots,\mathbf{C}_{m})},
\end{equation}
where the numerator is the marginal density of the multivariate exposure and the denominator is the mvGPS. These weights are referred to in the literature as stabilized inverse probability of treatment weights (IPTW) \citep{robins2000}. To motivate the intuition behind constructing weights in this manner, we can note that $w=1$ when the probability of exposure is independent of the confounding set $\C$, i.e., $f(\mathbf{D}\mid\mathbf{C}_{1},\dots,\mathbf{C}_{m})=f(\mathbf{D})$, which would hold in the case of a randomized experiment. For tractability, we propose using multivariate normal models for both densities, i.e., 
\begin{equation*}
    \mathbf{D}\sim \text{N}_{m}(\boldsymbol{\mu}, \boldsymbol{\Sigma}) \quad \mathbf{D}\mid \mathbf{C}_{1},\dots,\mathbf{C}_{m}\sim \text{N}_{m}\Bigg(\begin{bmatrix}\boldsymbol{\beta}_{1}^{T}\mathbf{C}_{1}\\ \vdots \\ \boldsymbol{\beta}_{m}^{T}\mathbf{C}_{m}\end{bmatrix}, \boldsymbol{\Omega}\Bigg),
\end{equation*} 
where each $\boldsymbol{\beta}^{T}_{j}$ is a row vector of length $p_{j}$ corresponding to the effect of the set of confounders $\mathbf{C}_{j}$ on $D_{j}$. By factorizing both the numerator and denominator in Equation~\ref{eqn:mvGPS_wts}, we can compute $w$ using full conditionals, i.e.,
\begin{equation}\label{eqn:mvGPS_cond_wts}
\begin{split}
    w=&\frac{f(D_{m}\mid D_{m-1},\dots, D_{1})\cdots f(D_{1})}{f(D_{m}\mid \mathbf{C}_{1},\dots,\mathbf{C}_{m},D_{m-1},\dots, D_{1})\cdots f(D_{1}\mid \mathbf{C}_{1},\dots,\mathbf{C}_{m})},\\
    w=&\frac{f(D_{m}\mid D_{m-1},\dots, D_{1})\cdots f(D_{1})}{f(D_{m}\mid \mathbf{C}_{m},D_{m-1},\dots, D_{1})\cdots f(D_{1}\mid \mathbf{C}_{1})},
\end{split}
\end{equation}
 where each conditional expression is univariate normal.  The second line is a result of the fact that the $j^{th}$ exposure is independent of the confounders of other exposures given $\mathbf{C}_{j}$, i.e.,
\begin{equation*}
    D_{j}\indp \mathbf{C}_{-j}\mid \mathbf{C}_{j}\ \forall \ j=1,\dots,m, 
\end{equation*}  
where $\mathbf{C}_{-j}$ represents the set of confounders excluding $\mathbf{C}_j$, i.e., $\mathbf{C}_{-j}= \{\mathbf{C}_{1},\dots,\mathbf{C}_{m}\}\smallsetminus \mathbf{C}_{j}$. Evaluating only the conditional densities reduces computational burden by eliminating the need to directly estimate the covariance matrices, $\boldsymbol{\Sigma}$ and $\boldsymbol{\Omega}$. 
 
 Let $\boldsymbol{\theta}$ be the collection of mean and variance parameters from all of the univariate normal densities in Equation~\ref{eqn:mvGPS_cond_wts}. Estimation of the parameters to obtain $\hat{\boldsymbol{\theta}}$ proceeds by maximizing the corresponding conditional density via least squares. The weight for the $i^{th}$ subject, $w_i$, is obtained by evaluating the densities using $\hat{\boldsymbol{\theta}}$ with the values of the observed exposures, $D_{i1}, \dots, D_{im}$, and confounders, $\mathbf{C}_{i1},\dots,\mathbf{C}_{im}$.

When the weights are properly specified, the covariance between each exposure $D_{j}$ and confounder $C_{jk}$ for $j=1,\dots,m$ and $k=1,\dots,p_{j}$ is zero:
\begin{equation}
    \begin{split}
        &\E[w(D_{j}-\mu_{D_j})(C_{jk}-\mu_{C_{jk}})]=\int_{\mathcal{D}}\int_{\mathcal{C}_{1}}\cdots\int_{\mathcal{C}_{m}}w(d_{j}-\mu_{D_{j}})(c_{jk}-\mu_{C_{jk}})f(\mathbf{d}, \mathbf{c}_{1},\dots,\mathbf{c}_{m})\partial \mathbf{d}\partial\mathbf{c}_{1},\dots,\partial\mathbf{c}_{m}\\
        &=\int_{\mathcal{D}}\int_{\mathcal{C}}\frac{f(\mathbf{d})}{f(\mathbf{d}|\mathbf{c}_{1},\dots,\mathbf{c}_{m})}(d_{j}-\mu_{D_{j}})(c_{jk}-\mu_{C_{jk}})f(\mathbf{d}, \mathbf{c}_{1},\dots,\mathbf{c}_{m})\partial \mathbf{d}\partial\mathbf{c}_{1},\dots,\partial\mathbf{c}_{m}\\
        &=\int_{\mathcal{D}}\int_{\mathcal{C}}\frac{f(\mathbf{d})f(\mathbf{c}_{1},\dots,\mathbf{c}_{m})}{f(\mathbf{d}|\mathbf{c}_{1},\dots,\mathbf{c}_{m})f(\mathbf{c}_{1},\dots,\mathbf{c}_{m})}(d_{j}-\mu_{D_{j}})(c_{jk}-\mu_{C_{jk}})f(\mathbf{d}, \mathbf{c}_{1},\dots,\mathbf{c}_{m})\partial \mathbf{d}\partial\mathbf{c}_{1},\dots,\partial\mathbf{c}_{m}\\
        &=\int_{\mathcal{D}}\int_{\mathcal{C}}(d_{j}-\mu_{D_{j}})(c_{jk}-\mu_{C_{jk}})f(\mathbf{d})f(\mathbf{c}_{1},\dots,\mathbf{c}_{m})\partial \mathbf{d}\partial\mathbf{c}_{1},\dots,\partial\mathbf{c}_{m}\\
        &=\int_{\mathcal{D}}(d_{j}-\mu_{D_{j}})f(\mathbf{d})\partial\mathbf{d}\int_{\mathcal{C}}(c_{jk}-\mu_{C_{jk}})f(\mathbf{c}_{1},\dots,\mathbf{c}_{m})\partial\mathbf{c}_{1},\dots,\partial\mathbf{c}_{m}\\
        &=0.
    \end{split}
\end{equation}
This balancing property of the weights serves as an important diagnostic when using the mvGPS as part of a causal analysis \citep{austin_gps_bal}. Weights that do not reduce the exposure-confounder correlation suggest that the distributional assumptions are invalid, the propensity equations are misspecified, or that there are insufficient data as the balance is achieved asymptotically.

Further, it follows that these weights are already normalized, i.e, $\E[w]=1$, and they maintain the marginal moments of $\mathbf{D}$ and $\mathbf{C}_{j}$, meaning $\E[w D_{j}]=\E[D_{j}]$ and $\E[w\mathbf{C}_{j}]=E[\mathbf{C}_{j}]$ for $j=1,\dots,m$, where the expectations are taken with respect to the joint density $f(\mathbf{D},\mathbf{C}_{1},\dots,\mathbf{C}_{m})$.

It remains to show that the weights as constructed satisfy Equation~\ref{eqn:unbias_w}. To do this we follow the logic proposed by Robins on using IPTW to correct for confounding \citep{robins2000marginal}. We first note that the joint density of the potential outcome can be factorized as
\begin{equation*}
    \begin{split}
    f(Y(\mathbf{d}), \mathbf{D}, \mathbf{C}_{1},\dots,\mathbf{C}_{m})&=f(\mathbf{D}\mid Y(\mathbf{d}), \mathbf{C}_{1},\dots,\mathbf{C}_{m})f(\mathbf{C}_{1},\dots,\mathbf{C}_{m}\mid Y(\mathbf{d}))f(Y(\mathbf{d}))\\
    &=f(\mathbf{D}\mid \mathbf{C}_{1},\dots,\mathbf{C}_{m})f(\mathbf{C}_{1},\dots,\mathbf{C}_{m}\mid Y(\mathbf{d}))f(Y(\mathbf{d})),
    \end{split}
\end{equation*}
where the second line follows from the assumption of weak ignorability. We can then let $f(\mathbf{D})$ be a density for our multivariate exposure and construct a new joint density $f^*$ where we replace $f(\mathbf{D}\mid\mathbf{C}_{1},\dots,\mathbf{C}_{m})$ with $f(\mathbf{D})$ as would be the case if the exposures were independent of the confounders. This new density is written as
\begin{equation*}
    f^{*}(Y(\mathbf{d}), \mathbf{D}, \mathbf{C}_{1},\dots,\mathbf{C}_{m})=f(\mathbf{D})f(\mathbf{C}_{1},\dots,\mathbf{C}_{m}\mid Y(\mathbf{d}))f(Y(\mathbf{d})),
\end{equation*}
where the marginal mean of the potential outcomes is equivalent under either joint density $f$ or $f^{*}$, i.e., ${\E^{*}[Y(\mathbf{d})]=\E[Y(\mathbf{d})]}$. Using this new density we can write our dose response as
\begin{equation*}
\mathbb{E}^{*}[Y(\mathbf{d})]=\mathbb{E}^{*}[Y(\mathbf{d})\mid\mathbf{D}=\mathbf{d}]=\mathbb{E}^{*}[Y(\mathbf{D})\mid \mathbf{D}=\mathbf{d}]=\mathbb{E}^{*}[Y\mid \mathbf{D}=\mathbf{d}],
\end{equation*} using the SUTVA assumption. The resulting expression, $\E^{*}[Y\mid \mathbf{D}=\mathbf{d}]$, is equivalent to the mean expression in a linear regression of the observed exposures on outcome. Finally, we have
\begin{equation}
\begin{split}
    \mathbb{E}^{*}[Y\mid \mathbf{D}=\mathbf{d}]&=\int_{\mathcal{Y}}\int_{\mathcal{D}}\int_{\mathcal{C}} y f^{*}(y,\mathbf{d}, \mathbf{c}_{1},\dots,\mathbf{c}_{m}) \partial y \partial \mathbf{d} \partial \mathbf{c}_{1},\dots,\partial\mathbf{c}_{m} \\
    &=\int_{\mathcal{Y}}\int_{\mathcal{D}}\int_{\mathcal{C}} y \frac{f^{*}(y,\mathbf{d}, \mathbf{c}_{1},\dots,\mathbf{c}_{m})}{f(y,\mathbf{d}, \mathbf{c}_{1},\dots,\mathbf{c}_{m})} f(y,\mathbf{d}, \mathbf{c}_{1},\dots,\mathbf{c}_{m}) \partial y \partial \mathbf{d} \partial \mathbf{c}_{1},\dots,\partial\mathbf{c}_{m}\\
    &=\int_{\mathcal{Y}}\int_{\mathcal{D}}\int_{\mathcal{C}} y \frac{f(\mathbf{d})}{f(\mathbf{d}\mid\mathbf{c}_{1},\dots,\mathbf{c}_{m})} f(y,\mathbf{d}, \mathbf{c}_{1},\dots,\mathbf{c}_{m}) \partial y \partial \mathbf{d} \partial \mathbf{c}_{1},\dots,\partial\mathbf{c}_{m}\\
    &=\int_{\mathcal{Y}}\int_{\mathcal{D}}\int_{\mathcal{C}} w y f(y,\mathbf{d}, \mathbf{c}_{1},\dots,\mathbf{c}_{m}) \partial y \partial \mathbf{d} \partial \mathbf{c}_{1},\dots,\partial\mathbf{c}_{m}\\
    &=\E[wY\mid\mathbf{D}=\mathbf{d}],
\end{split}
\end{equation}
which gives us the result from Equation~\ref{eqn:unbias_w} that our weighted regression does indeed provide a consistent estimate of the dose-response function.

\section{Simulation}\label{sim_section}
\subsection{Design}\label{sim_setup}
We conducted a simulation study to demonstrate the performance of the mvGPS method under different scenarios of confounding and compare it to three commonly used univariate methods. The univariate methods were entropy balancing \citep{tbbicke2020entropy}, the covariate balanced generalized propensity score (CBGPS) \citep{fong2018}, and the generalized linear propensity score (PS). The entropy balancing method uses non-parametric constrained optimization with an entropy loss function to solve for weights without specifying a propensity score model. CBGPS attempts to achieve propensity specification and covariate balance simultaneously by introducing a penalty term into the likelihood. The PS method uses univariate normal densities for the marginal distribution of exposure and the generalized propensity score without balance constraints.  Although these univariate methods can handle only single exposure variables, we expected that they might perform adequately when the multiple exposure variables are highly correlated and have the same confounders. However, when exposure variables have separate sets of confounders and/or are only weakly correlated, we expected that the mvGPS method would outperform the univariate methods. 

In our simulations we focus exclusively on a bivariate exposure, $m=2$, similar to that found in our motivating example. For each simulated data scenario, each univariate method was applied twice, once to each exposure variable, with each such application yielding a set of weights that were used to assess balance on confounders and estimate the dose-response function.  

The first step of the simulation is to draw the vector of covariates $\mathbf{X}$ for each unit. We assume that there are a total of $10$ covariates collected prior to exposure and that the covariates follow a normal distribution, \begin{equation*}
\mathbf{X}\sim \mathnormal{N}_{10}(\mathbf{0},\boldsymbol{\Sigma}_{X}),
\end{equation*}
where the covariance matrix $\boldsymbol{\Sigma}_{X}$ is compound symmetric with variance 1 and covariance 0.2, to create a set of correlated covariates. 

Realizations of the conditional distribution of the bivariate continuous exposure levels, $\mathbf{D}=(D_{1}, D_{2})^{T}$ given $\mathbf{X}$, were then generated as bivariate normal,
\begin{equation*}
\mathbf{D}\mid\mathbf{X}\sim \mathnormal{N}_{2}(\boldsymbol{\beta}\mathbf{X}, \boldsymbol{\Sigma}_{D\mid X}),
\end{equation*}
where $\boldsymbol{\beta}=\begin{bmatrix}\boldsymbol{\beta}^{T}_{1}\\ \boldsymbol{\beta}^{T}_{2}\end{bmatrix}$ is a $2\times10$ matrix with row vectors $\boldsymbol{\beta}^{T}_{1}$ and $\boldsymbol{\beta}^{T}_{2}$ representing the effects of $\mathbf{X}$ on $D_{1}$ and $D_{2}$, respectively, and $\boldsymbol{\Sigma}_{D\mid X}$ is the $2 \times 2$ conditional covariance matrix. For all simulations the conditional standard deviation for each exposure was set to 2, while values of the conditional correlation $\rho_{D\mid X}$ were allowed to vary over $\{0, 0.1, 0.3, 0.5, 0.7, 0.9\}$. 

Note that the marginal covariance matrix of the exposures, $\boldsymbol{\Sigma}_{D}$, is equal to $\boldsymbol{\Sigma}_{D}= \boldsymbol{\Sigma}_{D\mid X}+\boldsymbol{\beta}\boldsymbol{\Sigma}_{X}\boldsymbol{\beta}^{T}$. This means that the marginal correlation of the two exposure variables, $\rho_{D}$, depends on their conditional correlation $\rho_{D\mid X}$, the covariance of $\mathbf{X}$ and the degree of overlap of covariates. The degree of overlap is reflected in the number of non-zero elements that are common between $\boldsymbol{\beta}_{1}^{T}$ and $\boldsymbol{\beta}_{2}^{T}$. As the degree of overlap increases, the marginal correlation also increases. Since $\boldsymbol{\Sigma}_{X}$ is compound symmetric with constant covariance of $0.2$, the marginal correlation is guaranteed to be non-zero even with zero overlap and zero conditional correlation. 

Finally, the outcome $Y$ was sampled from a univariate normal distribution conditional on $\mathbf{D}$ and $\mathbf{X}$ as
\begin{equation*}
Y\mid\mathbf{D},\mathbf{X}\sim \mathnormal{N}\bigg(\boldsymbol{\alpha}^{T}\begin{bmatrix}\mathbf{X} \\ \mathbf{D}\end{bmatrix}, \sigma^{2}_{Y}\bigg)=\mathnormal{N}(\boldsymbol{\alpha}_{\mathbf{X}}^{T}\mathbf{X}+\boldsymbol{\alpha}_{\mathbf{D}}^{T}\mathbf{D}, \sigma^{2}_{Y}),
\end{equation*}
where $\boldsymbol{\alpha}^{T}=\begin{bmatrix}\boldsymbol{\alpha}_{\mathbf{X}}^{T} & \boldsymbol{\alpha}_{\mathbf{D}}^{T}\end{bmatrix}$ is a $1\times 12$ vector of coefficients, which we separate as $\boldsymbol{\alpha}_{\mathbf{X}}^{T}$, a $1\times 10$ vector representing the effect of covariates on the outcome, and  $\boldsymbol{\alpha}_{\mathbf{D}}^{T}$, a $1\times 2$ vector corresponding to the treatment effects. In all simulations the conditional standard deviation of the outcome was equal to 4, i.e., $\sigma_{Y}=4$.

\begin{figure}[ht]
\centering
\caption{Simulation Scenarios}
\includegraphics[height = 0.8\textheight, keepaspectratio]{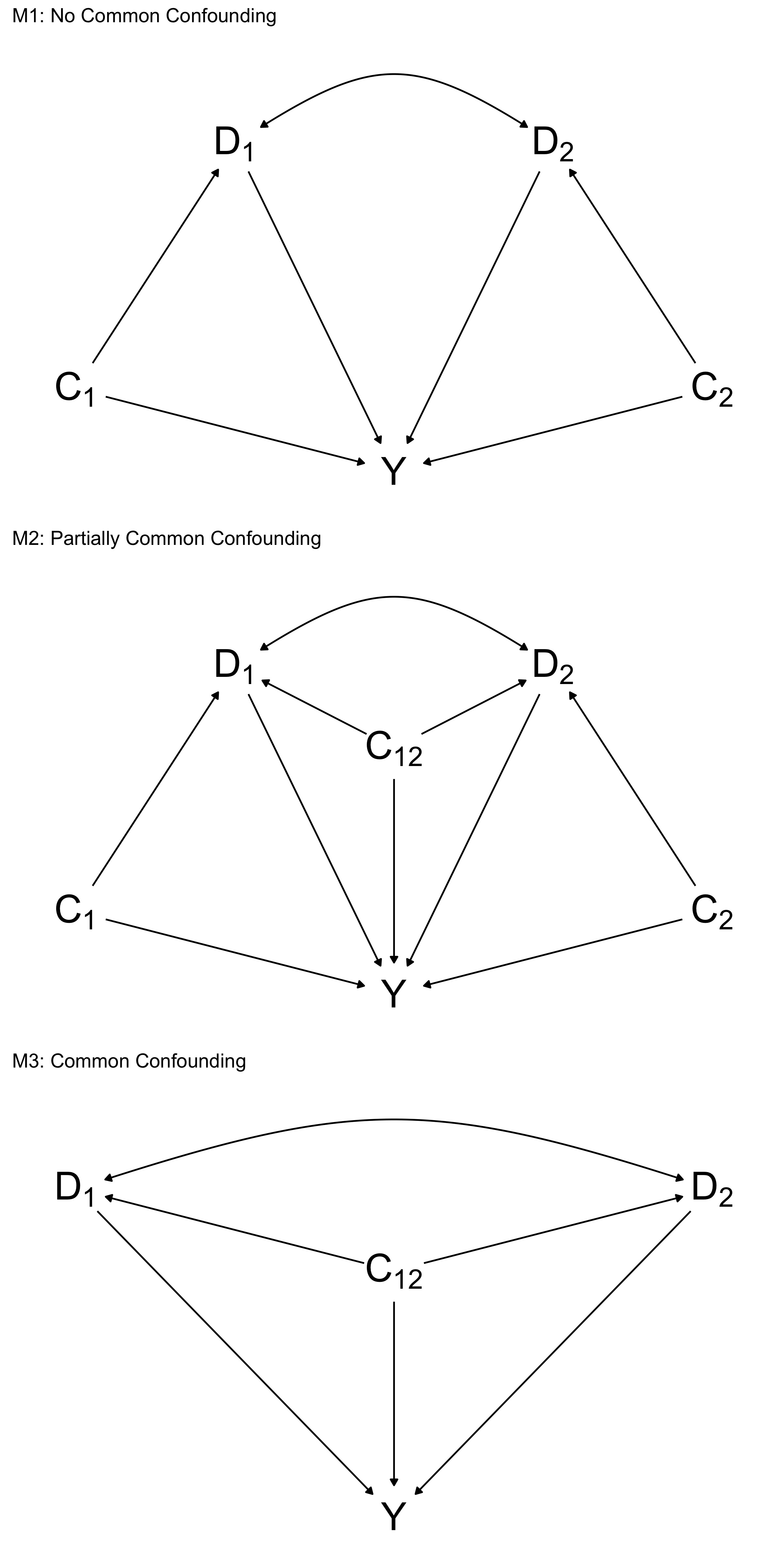}
\caption*{Directed acyclic graphs for each of the three data generating simulation scenarios. $D_{1}$ and $D_{2}$ are continuous exposure measures and $Y$ is the outcome of interest. $C_{1}$ and $C_{2}$ represent confounder sets that are specific to exposures $D_{1}$ and $D_{2}$, while $C_{12}$ represents a confounder set common to both exposures.}
\label{fig:simdag_scenarios}
\end{figure}

Three scenarios were constructed to reflect different degrees of overlap of confounding for the two exposures: M1: No Common Confounding, M2: Partially Common Confounding, and M3: Common Confounding. Directed acyclic graphs (DAGs) for each scenario are shown in Figure~\ref{fig:simdag_scenarios}. 

Tables~\ref{tab:m1_table},~\ref{tab:m2_table},~and~\ref{tab:m3_table} display the coefficients in the vectors $\boldsymbol{\beta}_1^{T}$, $\boldsymbol{\beta}_2^{T}$ and $\boldsymbol{\alpha}^{T}$ for each scenario. In M1, the two exposures $D_1$ and $D_2$ each have five covariates, with none in common; for each exposure, two of the covariates are true confounders (i.e., also associated with $Y$). The outcome $Y$ is a function of the two exposures as well as the four true confounders, none of which are shared between $D_1$ and $D_2$. In M2, $D_1$ and $D_2$ again have five covariates each, but they share three in common.  Two of the shared covariates are true confounders, and each exposure has a confounder that is not shared with the other exposure. The outcome $Y$ is again a function of the two exposures and the four true confounders, two of which are shared and two of which are not. In M3, $D_1$ and $D_2$ share the same five covariates. Four of these are true confounders, and $Y$ is a function of the two exposures and four common confounders. In all scenarios, the treatment effect for each of the exposures was set to 1, i.e., $\boldsymbol{\alpha}_{\mathbf{D}}^{T} = (1,1)$.

The three simulation scenarios were run with a sample size of $n=200$ for a total of $B=1000$ Monte Carlo repetitions using R Version 4.0 \citep{R_ref}. For each repetition, weights were estimated using mvGPS and the three univariate methods with the proper set of confounders specified for each exposure.  For example, for Scenario M1, weights were constructed for $D_1$ using $X_{2}$ and $X_{4}$ while $D_{2}$ depended on $X_{6}$ and $X_{9}$ (see Table~\ref{tab:m1_table}). The three univariate methods, entropy balancing, CBGPS and PS, were implemented using the \texttt{WeightIt} package in R \citep{weightIt}. The parametric version of CBGPS was used. The mvGPS method was implemented using the \href{https://github.com/williazo/mvGPS}{mvGPS} package in R. All methods were compared against an unweighted approach equivalent to applying a weight of 1 to all observations.

Weighted Pearson correlations between the exposures and confounders were used to assess balancing performance \citep{zhu_boosting,austin_gps_bal}. We examined maximum absolute correlation, which reflects the most imbalanced confounder after weighting and has been shown to be a key metric to assess balance \citep{diamond2013genetic}, and the average absolute correlation, which summarizes how well balance is achieved over all confounders. These correlation values were taken over both sets of exposures.  

Effective sample sizes, $(\Sigma_i w_i)^{2}/\Sigma_i w_i^2$, were calculated to summarize the relative power of each method \citep{kish_ess}. The weights were then used to estimate the dose-response model. The performance metrics were absolute total bias, $\Sigma_{j}|\alpha_{D_{j}}-\hat{\alpha}_{D_{j}}|$ for $j=1,2$, and root mean squared error, $\sqrt{\frac{1}{n}\Sigma_{i}(y^{*}_{i}-\hat{y}_i)^{2}}$, where $i=1, \dots, 500$ samples, $y^{*}$, were drawn from a uniform grid on the convex hull, $\mathcal{H}$, over the observed joint distribution of the two exposures. Each of the metrics was averaged over the $1000$ repetitions.

All methods reduce the effective sample size when weights are applied to the sample. Of particular concern for practitioners are extreme weights. When sample sizes are small or moderate, extreme weights can have an outsized influence. They may also result in limited power to detect treatment effects and erratic estimation \citep{kang2007demystifying}. One remedy is to trim extreme weights \citep{lee2011weight, huber2013performance}. As all simulations were run with moderate sample sizes, i.e., $n=200$, we wanted to test performance when weight trimming was applied as might be done in practice by analysts when faced with extreme weights. Our simulation analyses were thus repeated using trimmed weights for each method, $w_{q}$, where $q\in\{0.99, 0.95\}$. Weights were trimmed at both the upper and lower bounds of the respective sample percentile such that values above or below the thresholds were replaced with the threshold value.

\subsection{Simulation Results}\label{sim_results}

\begin{figure}[ht]
\centering
\caption{Assessing Covariate Balance: Maximum Absolute Exposure-Covariate Correlation}
\includegraphics[width = \textwidth, keepaspectratio]{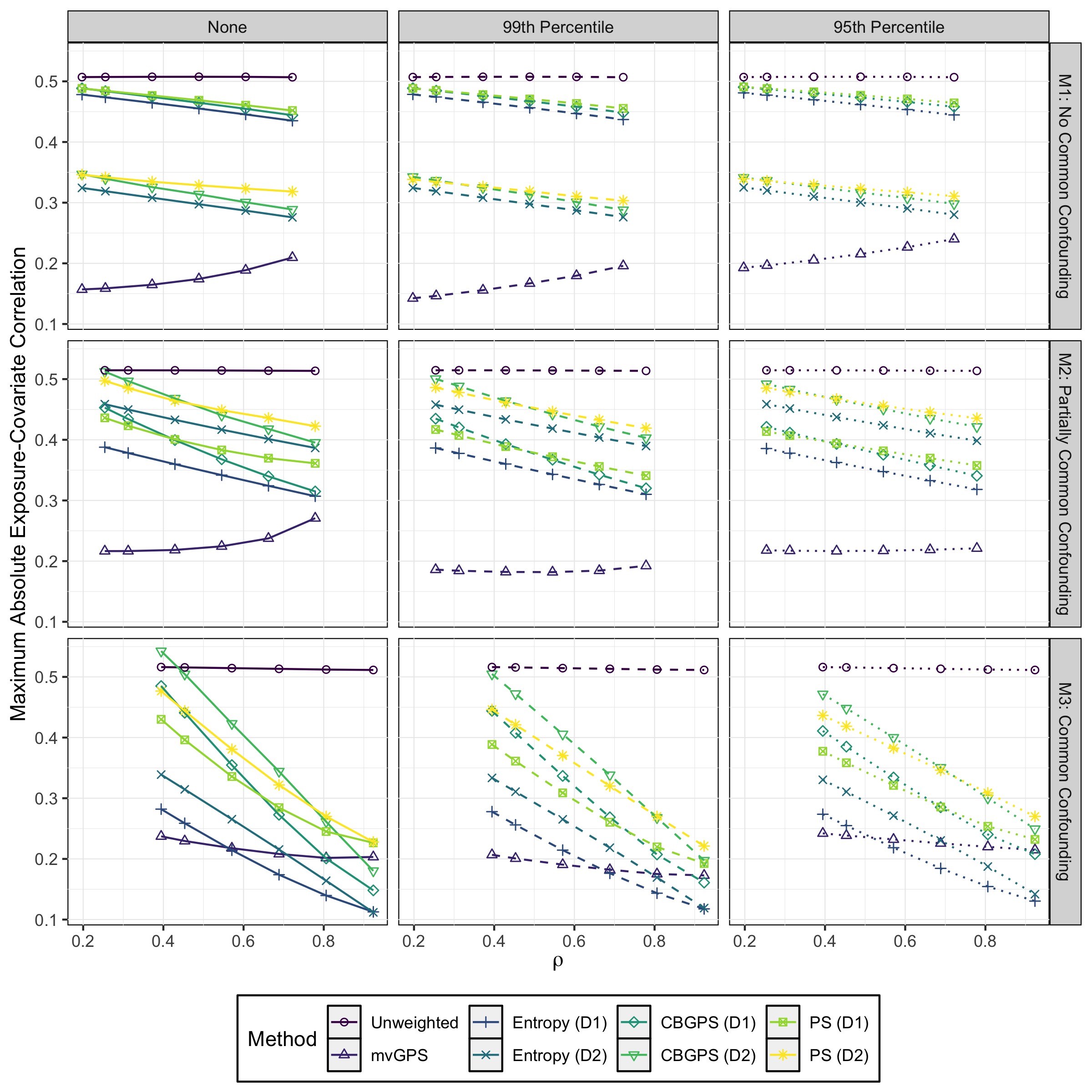}
\caption*{Rows correspond to the three simulation scenarios, M1, M2 and M3, and each column corresponds to quantiles used for weight trimming. The y-axis is the average maximum absolute exposure-covariate correlation for $n=200$ from $B=1000$ repetitions. This maximum is taken across both exposure values, $D_{1}$ and $D_{2}$. The x-axis, $\rho$, is the marginal correlation of the exposures. For univariate methods, weights were generated twice, once for each exposure variable.}
\label{fig:cov_bal_max_corr}
\end{figure}

Figure~\ref{fig:cov_bal_max_corr} plots the absolute maximum correlation between the exposures and confounders for each method along with the original unweighted correlations for comparison. In general, with no common confounding or partially common confounding, the mvGPS method substantially outperformed each univariate method. However, for common founding, mvGPS performs best only when the marginal correlation is low. The performance of univariate methods tended to cluster differently based on the degree of confounding overlap. In models with low overlap, performance was clustered based on exposure, $D_{1}$ or $D_{2}$, but as the degree of overlap increased, performance became clustered by type of estimation, Entropy, CBGPS, or PS. Applying trimmed weights, we see a slight improvement for $q=0.99$ while $q=0.95$ has little to no effect.

\begin{figure}[ht]
\centering
\caption{Assessing Covariate Balance: Average Absolute Exposure-Covariate Correlation}
\includegraphics[width = \textwidth, keepaspectratio]{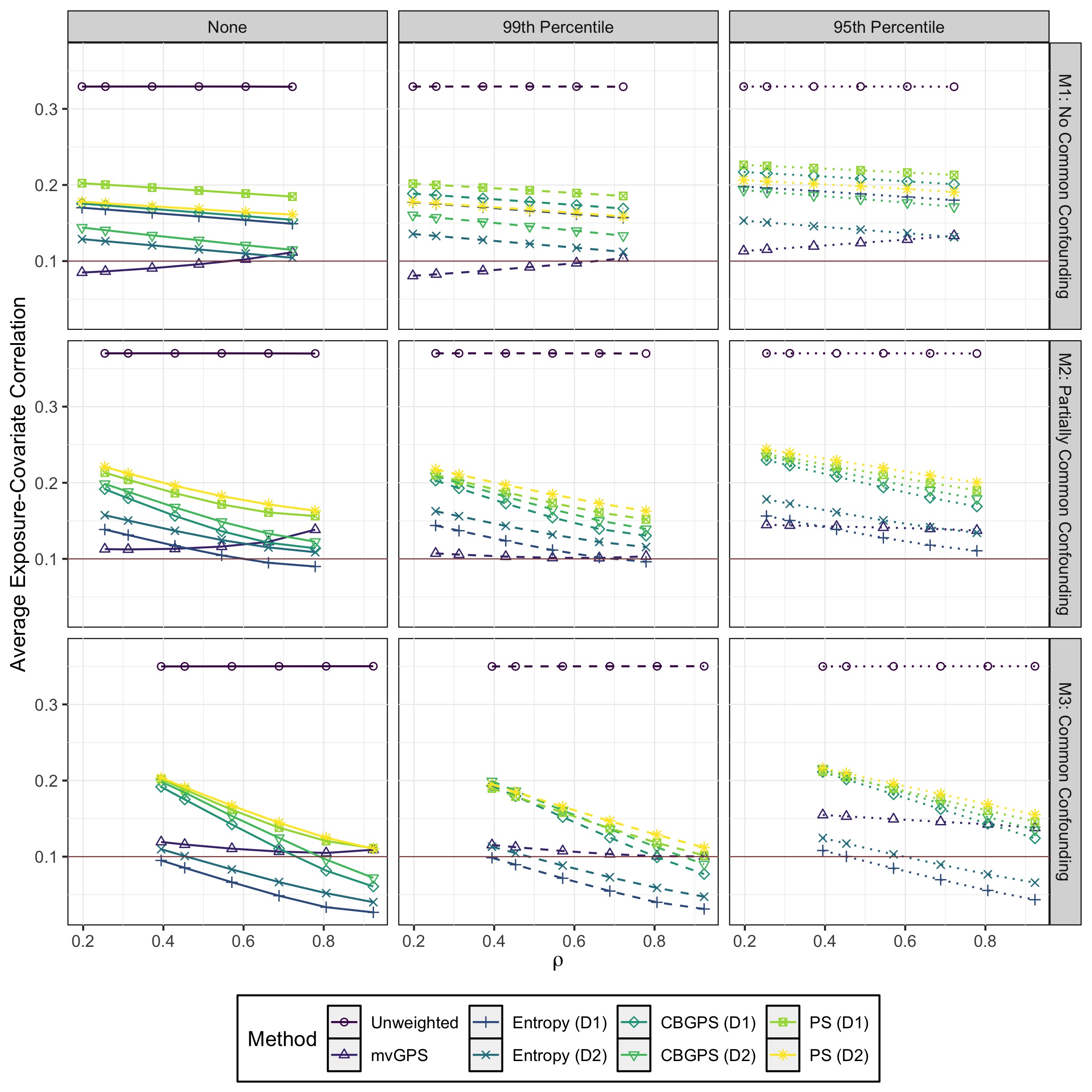}
\caption*{Rows correspond to the three simulation scenarios, M1, M2 and M3, and each column corresponds to quantiles used for weight trimming.  The y-axis is the average absolute exposure-covariate correlation for $n=200$ from $B=1000$ repetitions. This average is taken across both exposure values, $D_{1}$ and $D_{2}$. The x-axis, $\rho$, is the marginal correlation of the exposures. For univariate methods, weights were generated twice, once for each exposure variable. The \textcolor{indianred4}{red} line corresponds to an average value of 0.1, which is often used as a benchmark for sufficient balance.}
\label{fig:cov_bal_avg_corr}
\end{figure}

Figure~\ref{fig:cov_bal_avg_corr} shows the average absolute exposure-covariate correlation along with a reference line at $0.1$, a benchmark suggesting sufficient covariate balance \citep{zhu_boosting}. For all simulation models, the mvGPS is consistently near the $0.1$ threshold. For the univariate methods, we see trends in performance similar to those observed for the maximum correlation, but differences between methods are smaller. Entropy methods consistently had the lowest average correlation, especially with high overlap or high marginal correlation. Trimming the weights tended to eliminate the effect of the marginal correlation on the mvGPS, resulting in flatter lines for $q=0.99$ and $q=0.95$, particularly for models with at least some common confounding.

\begin{figure}[ht]
\centering
\caption{Effective Sample Size}
\includegraphics[width = \textwidth, keepaspectratio]{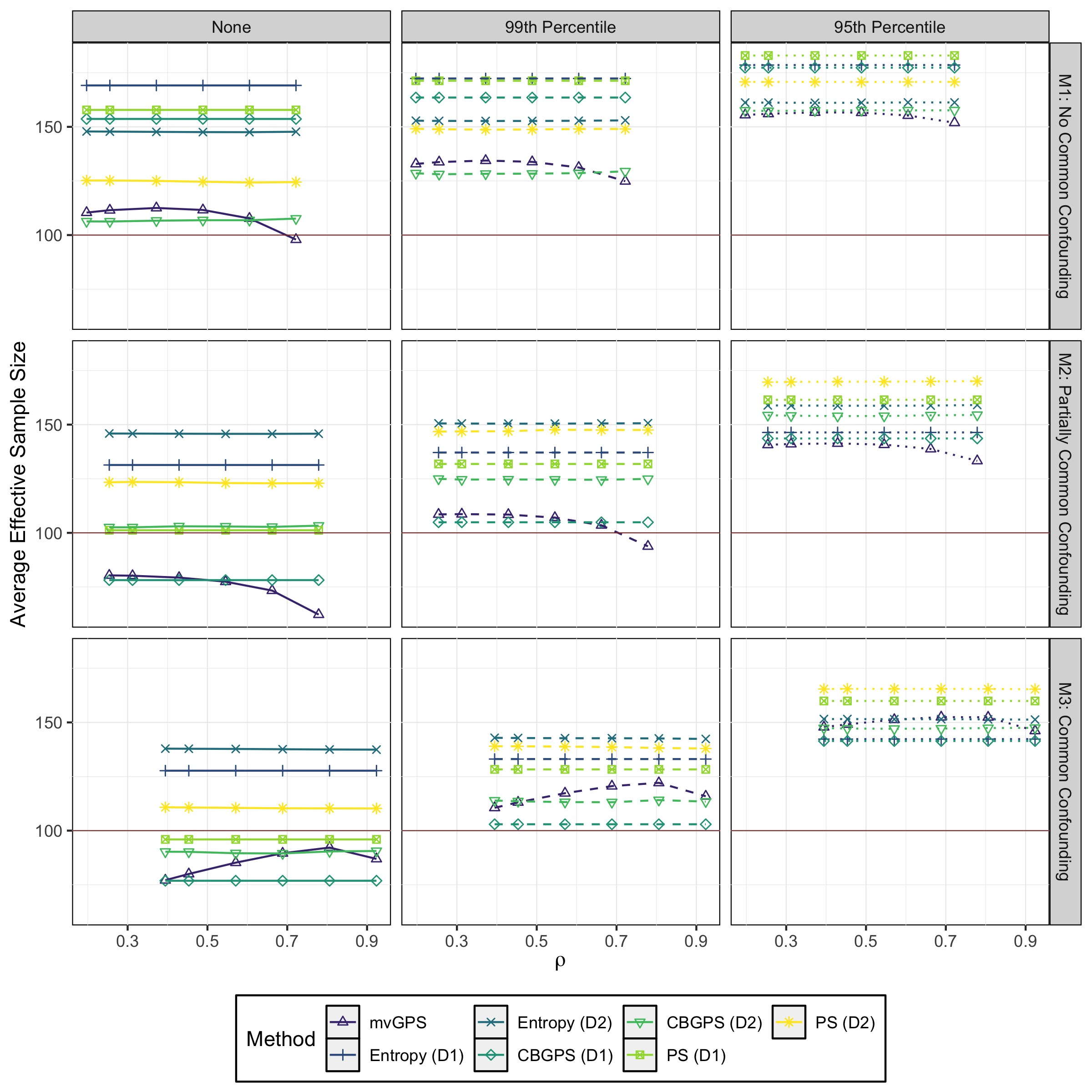}
\caption*{Rows correspond to the three simulation scenarios, M1, M2 and M3, and each column corresponds to quantiles used for weight trimming. The y-axis is the average effective sample size, $(\Sigma_i w_i)^2/\Sigma_i w_i^{2}$,  for $n=200$ from $B=1000$ repetitions. The x-axis, $\rho$, is the marginal correlation of the exposures. The \textcolor{indianred4}{red} line corresponds to an effective sample size of 100 which is often a minimum desirable quantity for hypothesis testing and inference of the dose-response model.}
\label{fig:sim_ess}
\end{figure}

Figure~\ref{fig:sim_ess} displays trends in effective sample size for each method across the various simulation scenarios, with a reference line at $100$, which is often a minimum desirable quantity for inference in the dose-response model \citep{vegetabile2020nonparametric}. The mvGPS method tends to have lower effective sample sizes compared to the univariate methods. Importantly, using untrimmed weights, the mvGPS method has effective sample sizes less than $100$ in the presence of partial or common confounding, indicating particularly low power. Trimming the weights increases the effective sample size for all methods and the difference between methods decreases as $q$ increases.

\begin{figure}[ht]
\centering
\caption{Outcome Modeling Performance Metric: Average Total Absolute Bias}
\includegraphics[width = \textwidth, keepaspectratio]{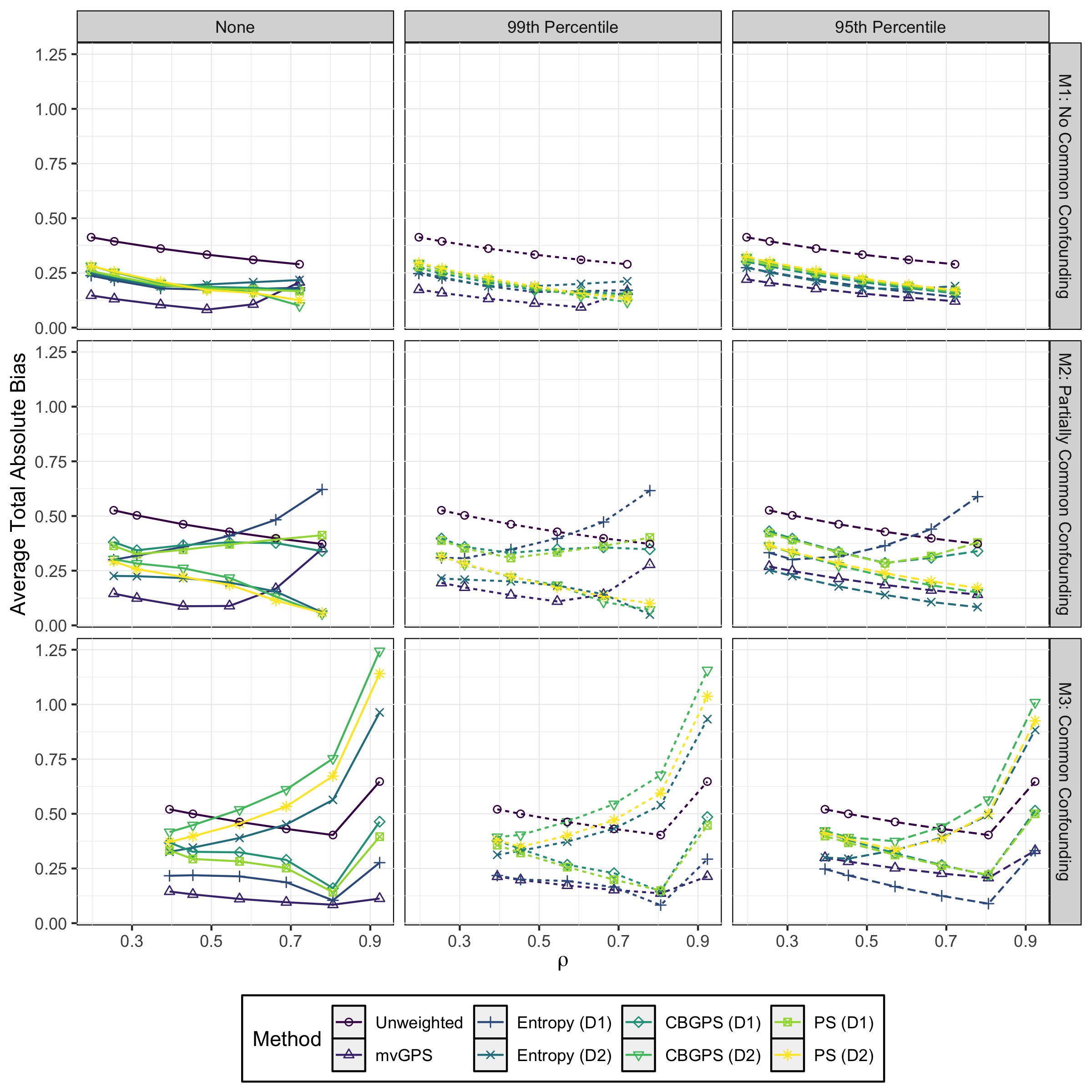}
\caption*{Rows correspond to the three simulation scenarios, M1, M2 and M3, and each column corresponds to quantiles used for weight trimming. The y-axis is the average total absolute bias, $\Sigma_j |\alpha_{D_j}-\hat{\alpha}_{D_{j}}|$,  for $n=200$ from $B=1000$ repetitions. The x-axis, $\rho$, is the marginal correlation of the exposures. For univariate methods, weights were generated twice, once for each exposure variable. }
\label{fig:sim_bias}
\end{figure}

Figure~\ref{fig:sim_bias} shows the results of each method with respect to total absolute bias for the treatment effects estimated from weighted regression. Generally, the mvGPS method has the lowest total bias, with the exception of high correlation in the model with partially common confounding or no common confounding. Of particular note, although the effective sample size and balancing diagnostics were lower for mvGPS in the common confounding model, it significantly outperforms all univariate methods with respect to bias even with high marginal correlation. We also observe that certain univariate methods have greater bias than the unweighted estimates, such as those that estimate weights using $D_{2}$ for the common confounding model. Trimming the weights tended to reduce bias for mvGPS when there was high correlation of the exposures, particularly under models with either partially overlapping confounding or no common confounding, while slightly increasing the bias for the common confounding models.

\begin{figure}[ht]
\centering
\caption{Outcome Modeling Performance Metric: Average RMSE}
\includegraphics[width = \textwidth, keepaspectratio]{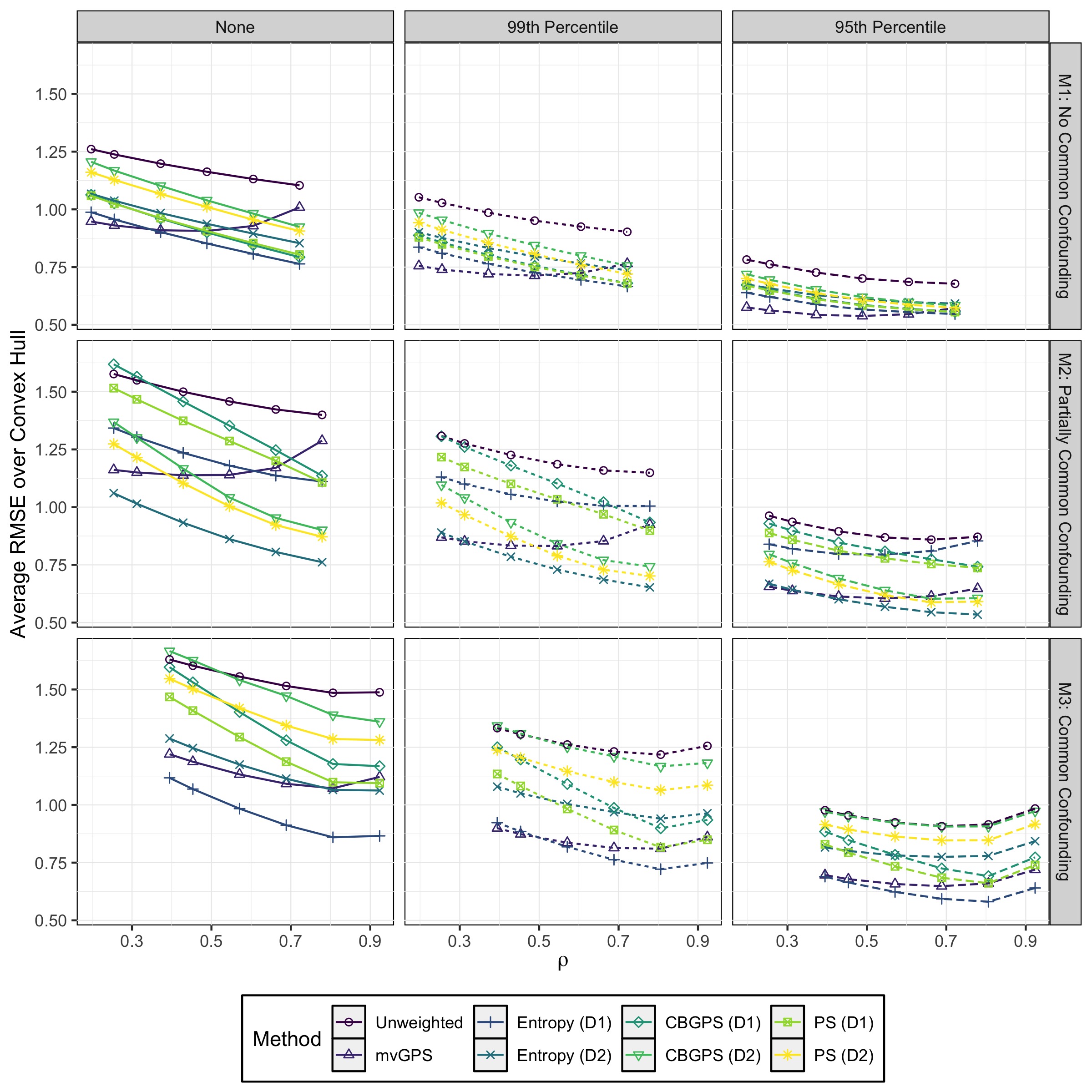}
\caption*{Rows correspond to the three simulation scenarios, M1, M2 and M3, and each column corresponds to quantiles used for weight trimming. The y-axis is the average root mean squared error (RMSE) for $500$ points sampled on a convex hull $H_q$ grid for $n=200$ from $B=1000$ repetitions. The x-axis, $\rho$, is the marginal correlation of the exposures. For univariate methods, weights were generated twice, once for each exposure variable. }
\label{fig:sim_hull_mse}
\end{figure}

Figure~\ref{fig:sim_hull_mse} shows the root mean squared error based on $500$ points sampled along a grid from the convex hull, $\mathcal{H}_{q}$, of the exposures. The precision of predicted values for the mvGPS is often worse than that for univariate methods. As $\rho$ increases, the mvGPS method has worse performance, with decreased power from low effective sample sizes. Trimming the weights helps reduce this trend and reduces the root mean squared error across all methods.

In summary, using a multivariate method for weight estimation is critical to achieve balance as univariate methods in general do not effectively balance on the confounders for both exposures. The multivariate method protects against any single confounder being strongly imbalanced across either exposure at the expense of slightly lower average balance, while the univariate methods have potentially large imbalance on the unused exposure dimension. The notable exception is when there is high overlap in terms of confounders. In this case univariate methods can sufficiently balance confounders, particularly when the marginal correlation of exposures is high. However, despite achieving balance in these scenarios, the univariate methods still resulted in high total bias of the treatment effect estimates. Although mvGPS weights were advantageous with respect to balance and bias, they tend to produce smaller effective sample sizes, resulting in lower power and higher root mean squared error. Weight trimming, particularly with $q=0.99$, offers a potential remedy to reduce these effects while also maintaining balance and low bias.

\section{Effect of Intervention Strategies on Childhood Obesity Prevalence}\label{wic_results}
The WIC program is designed to provide nutrition education, 'vouchers' for selected healthy food, and referrals. The rapid increase in childhood obesity rates in the early 2000s led to interventions to increase physical activity, and improve access to healthy food especially in communities where affordable fresh produce is not available. In this motivating example, we estimated the causal effects of interventions that were implemented by individual WIC clinics in attempts to meet the specific needs of the communities they served. As discussed in Section~\ref{background}, the intervention programs were classified as using macro or micro strategies and we calculated two continuous dose measures for $n=1079$ census tracts. The outcome was the difference in average obesity prevalence from post, 2012-2016, to pre, 2007-2009, intervention period, calculated as $Y=\bar{p}_{post}-\bar{p}_{pre}$ at the census tract level. Negative values of $Y$ indicate that the prevalence of obesity decreased. We hypothesized that areas with more macro and micro strategies would have the greatest reduction in rates of obesity. 

Data on potential census tract-level confounders came from three sources: US Census American Community Survey (ACS) 5-year estimates \citep{census}, WIC administrative data, and the National Establishment Time-Series (NETS) \citep{walls2013national}. Variables from the ACS captured community level demographic characteristics such as median household income, education level, primary language spoken at home, and ethnicity, which have been shown in previous research to be associated with obesity rates \citep{nobari2013immigrant, Nobari2018aa}. WIC administrative data were used to calculate average pre-treatment overweight and obesity prevalence for each census tract. Overweight and obesity prevalence were considered potential confounders because agencies may have directed interventions towards clinics in higher prevalence regions. Finally, NETS provided information on neighborhood food environments, specifically on the density per square mile of unhealthy and healthy food outlets \citep{Wang2006aa, anderson2020neighborhood}. Previous analysis has shown that higher density of healthy outlets was associated with lower obesity prevalence among low-income preschool-aged children in Los Angeles County \citep{chaparro2014influences}. Both macro and micro propensity dose equations included the same set of potential confounders from these three data sources, but each exposure was assessed separately to determine if higher order polynomial terms for any confounders were needed. These analyses showed that both macro and micro dose had quadratic relationships with education level and density of food outlets, while only macro dose had evidence of a quadratic relationship with median household income. 

After defining the appropriate functional form for each exposure and confounder, weights were then estimated using the mvGPS method and the three univariate methods discussed in Section~\ref{sim_setup}. To maintain the assumption of positivity, data used for estimating the weights were restricted to the trimmed convex hull $H_{0.95}$ shown in Figure~\ref{fig:macro_micro_joint}, where a bivariate normal distribution is plausible. The marginal correlation of exposures was moderate, $r=0.28$, in this high-density region. As both exposures had nearly identical sets of potential confounders, the data generating mechanism was akin to the common confounding scenario described in the simulations. Therefore, to protect against extreme weights and reduce the variability of the resulting dose-response estimates, weights for each method were trimmed using $q=0.99$. 

Table~\ref{tab:cov_bal} shows the balancing diagnostics, maximum absolute correlation and average absolute correlation, and the effective sample sizes. The confounders were significantly imbalanced prior to weighting with the average absolute correlation above $0.2$ and the maximum absolute correlation above $0.4$. All methods were able to improve balance, but the mvGPS method had substantially greater reduction in imbalance than the univariate methods. The average absolute correlation and maximum absolute correlation were reduced to $0.04$ and $0.10$ respectively after applying mvGPS weights. The effective sample size for mvGPS was reduced from the original sample of $n=1079$ to $604$. However, since the population included over $1000$ census tract units, the power was still reasonably high.

Finally, we applied weights from the mvGPS method to estimate the joint effect of macro and micro exposure doses on change in obesity prevalence using weighted least squares regression and compare these to unweighted estimates. Only exposures within the trimmed convex hull, $H_{0.95}$, were used to estimate treatment effects and predict the dose-response surface. The dose-response model for both methods included linear terms for each exposure and an interaction between the two exposures.

\begin{figure}[ht]
\centering
\caption{Estimated Dose-Response Surface of Change in Obesity Prevalence as a Function of Macro and Micro Intervention Dose}
\includegraphics[width = \textwidth, keepaspectratio]{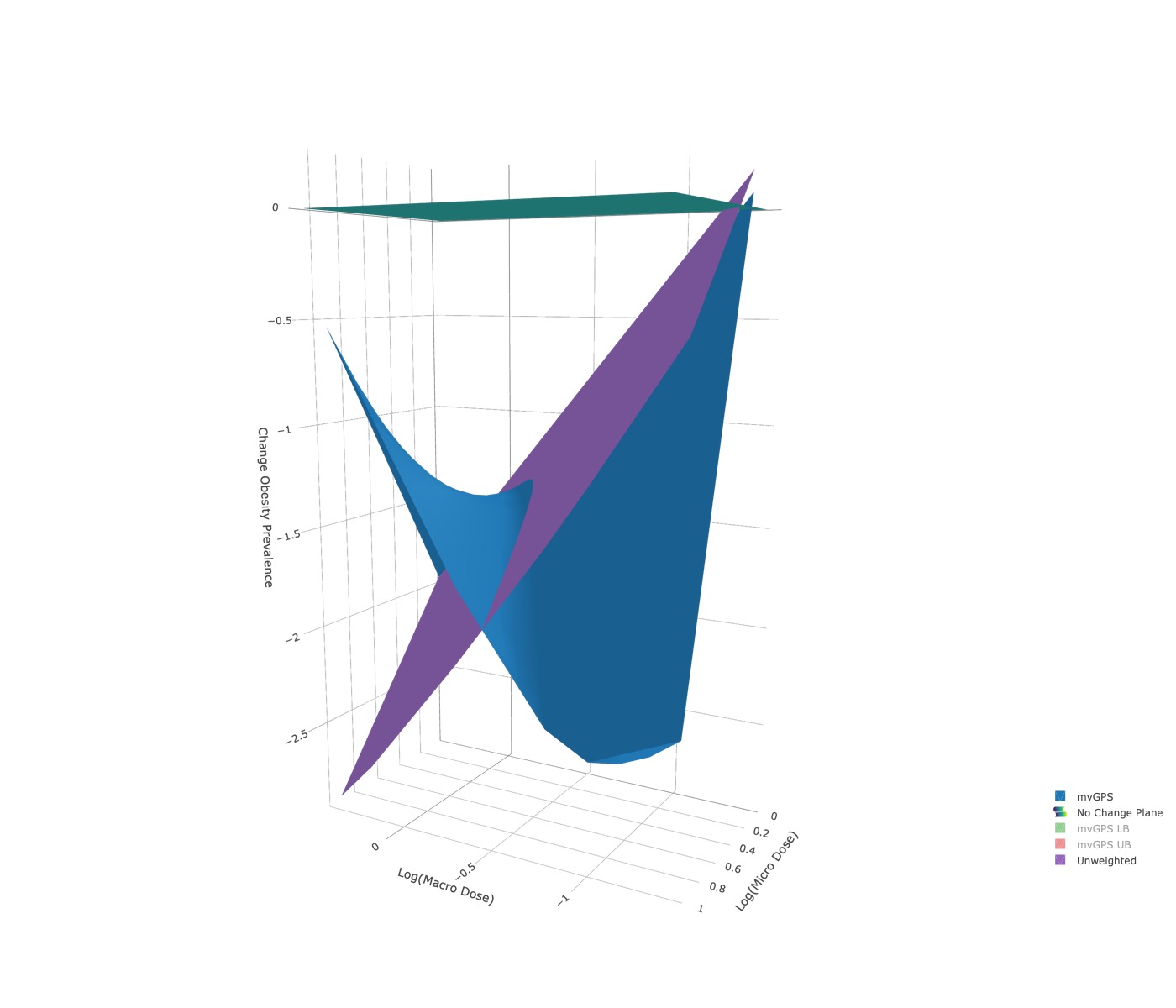}
\caption*{Estimated dose-response surface of change in obesity prevalence as a function of $\log$ macro and micro dose, obtained using mvGPS weights and unweighted. The surface is restricted to the convex hull of observed bivariate exposure, $H_{0.95}$, shown in Figure~\ref{fig:macro_micro_joint} and points are sampled evenly along this grid. A reference plane of no change is included. For a 3D interactive version of the dose-response surface that includes lower and upper bound 95\% confidence interval surfaces for the mvGPS method, visit \url{https://williazo.github.io/resources/}.}
\label{fig:dose_response}
\end{figure}

Figure~\ref{fig:dose_response} shows the weighted and unweighted dose-response surfaces along with a reference plane of no change in obesity prevalence. Both the weighted and unweighted surfaces suggest reductions in obesity prevalence from our pre-intervention period, 2007-2009, to the post-intervention period, 2012-2016, for all dose combinations. This is consistent with studies showing a decrease in obesity risk among WIC-participating children in Los Angeles County associated with the 2009 change in the WIC food packages \citep{Chaparro:2019aa, nobari2018trends}. The unweighted dose-response surface is a monotonic plane; increases in micro dose and in macro dose are each associated with greater reduction in obesity prevalence, the associations are additive, and the greatest reduction in prevalence corresponds to the highest levels of macro and micro doses. The mvGPS dose-response surface is more complex and shows an interaction effect. At low levels of macro dose, increases in micro dose are associated with a steep reduction in obesity prevalence. However, as macro dose increases, high micro dose becomes gradually less effective. In the quadrant where both macro and micro dose are high, higher micro doses appear to be less beneficial rather than more beneficial. There are several possible explanations for this result. There could be important confounders that were not accounted for in the analysis. There could also have been measurement error in estimating exposures. We noted that the data set included several observations with high macro and high micro doses that were assigned high weights and had either no decrease or a slight increase in child obesity prevalence. Further investigation of these census tracts may yield more information and guide model refinements.

\section{Discussion}\label{discussion}
In this manuscript, we introduced methodology for generating a multivariate generalized propensity score, mvGPS, to be used in estimating the causal effect of multiple simultaneous continuous exposures in observational or non-randomized studies. We have developed the R package \texttt{mvGPS} available at \url{https://github.com/williazo/mvGPS} to implement the methods.

Through simulations we have shown that, when estimating the causal effects of two simultaneous exposures, mvGPS weights are effective at reducing both the maximum and average absolute correlation between exposures and confounders. Further, the weights can minimize bias in estimating the dose-response function in realistic data generating settings with moderate sample sizes. The simulations identified two key factors that affect performance. When the exposures have highly overlapping sets of confounders or large marginal correlation, the mvGPS method may generate extreme weights, resulting in smaller effective sample sizes and higher average root mean squared error. Trimming the weights improves performance in these situations. We suggest that in settings with a high degree of overlap in the confounders or moderate to large marginal exposure correlation, weights should be trimmed at $q=0.99$.

While our method could in principle be extended to an arbitrary number of continuous exposures, we have confined attention in our simulations and application to the setting of two exposures. Assessing the joint effect of two interventions is a common scientific question, and we expect that there are many practical applications of the methods. The resulting dose-response surface for bivariate exposures can be easily visualized and interpreted, which is key in practice. Further work is needed to explore performance for settings with more than two exposures. In particular, positivity and achieving adequate balance may be increasingly difficult as dimensions of exposure increase. For higher order exposures, dimension reduction techniques such as principal component analysis or manifold learning might be applied to transform the problem to a lower order continuous exposure space.

We applied the mvGPS method to evaluate the joint effectiveness of macro and micro intervention strategies used in childhood obesity programs on change in obesity prevalence among low-income preschool aged children. Due to non-random selection of participating clinics, there was significant imbalance on potential confounders as evidenced by large absolute maximum and average correlations prior to weighting. The mvGPS method achieved superior balance compared to univariate alternatives, reducing the maximum absolute correlation to the benchmark of 0.1 and the average absolute correlation to nearly zero. Estimates of the dose-response surface using weights from the mvGPS method differed substantially from the results of the unweighted surface. The results showed that the most effective intervention combination was higher levels of micro strategies and lower levels of macro strategies. However, our results should be interpreted with caution. As with other causal inference methods, all confounders must be adequately captured and modeled to produce unbiased estimates. Thus our estimated treatment effects may be biased due to unknown confounders. In particular, communities that received higher levels of macro and micro doses may have been inherently more difficult to change due to a complex interplay of community and personal factors not captured by our set of potential confounders. In addition, we have assumed that the potential outcomes of the change in obesity given macro and micro exposures are well-defined via SUTVA as discussed in Section~\ref{subsec:id_assump}. Specifically, SUTVA stipulates that multiple versions of the exposures do not exist. In our application, however, there are potentially different sets of interventions that can yield the same macro and micro exposure scores. In the presence of multiple versions of the exposures, the resulting potential outcomes may be unidentifiable. Further, we had to estimate exposure to the interventions at the census tract level using various assumptions, which may have resulted in measurement error. Thus our application, while demonstrating the methods, has important limitations. 

Our approach assumes that the exposures have a multivariate normal distribution. The multivariate normal distribution is particularly attractive for working in higher dimensions. In our case, it allows for the full conditionals used to generate weights in Equation~\ref{eqn:mvGPS_cond_wts} to be tractable univariate normal densities. Further, the asymptotics of the estimates are well behaved due to the central limit theorem. We note that it is common practice when using the generalized propensity score for continuous treatments in the univariate case to assume normality \citep{hirano_continuous, imai_causalGPS, zhu_boosting, robins2000marginal, fong2018}. However, this reliance on multivariate normality is an important limitation of the methodology, particularly in assessing its validity \citep{mecklin2004}. Possible extensions include replacing the multivariate normal distribution with non-parametric or semi-parametric alternatives as has been done recently with univariate methods \citep{kennedy2017, tbbicke2020entropy, vegetabile2020nonparametric}.  Other possible extensions include allowing for time-varying outcomes and exposures such has been recently proposed for CBGPS \citep{CBGPS_time_varying}.  Additionally, SUTVA might be relaxed to test for potential geographic interference \citep{vanderweele2008, verbitsky2012causal}. 

\section*{Acknowledgments}
This work was supported in part by funding from the National Institute of Health R01 HD072296. We would like to thank May Wang and Michael Prelip, co-principal investigators of the study, for their feedback and help throughout the development of this project. Special thanks to Shannon Whaley and Christopher Anderson from PHFE WIC in providing access to the data from WIC and helping to generate catchment areas. Finally, thanks to the members of the ECOSyS team that have provided assistance on this project: Shelley Jung, Linghui Jiang, Roch Nianogo, Tabashir Nobari, Kara MacLeod, and Lilly Nhan. The opinions about the data used in the manuscript are solely those of the authors and do not reflect official positions of either government or individual agencies.

\clearpage

\bibliographystyle{DeGruyter}
\bibliography{mvGPS}

\clearpage
\section*{Tables and Figures}
\begin{table}[ht]
    \centering
    \caption{Intervention Program Strategies}
    \begin{tabular}{|c|c|c|c|}
        \hline
        \textbf{\#} & \textbf{Name} & \textbf{Group} & n (\%)\\
        \hline
        1 & Government Policies & Macro & 1 (3\%)\\
        2 & Institutional Polices & Macro & 4 (12\%)\\
        3 & Infrastructure Investments & Macro & 3 (9\%)\\
        4 & Business Practices & Macro & 4 (12\%)\\
        \hline
        5 & Group Education & Micro & 21 (66\%)\\
        6 & Counseling & Micro & 14 (44\%)\\
        7 & Health Communication & Micro & 17 (53\%)\\
        8 & Home Visitation & Micro & 8 (25\%)\\
        9 & Screening and Referral & Micro & 14 (44\%)\\
        \hline
    \end{tabular}
    \caption*{Each intervention program was classified based on the strategies that were implemented. Programs could use multiple strategies, e.g., Group Education and Counseling. Strategies are categorized as either micro and macro based on whether they directly targeted individuals or the population at large. The final column represents how many of the 32 programs used that particular strategy.}
    \label{tab:strat_tbl}
\end{table}

\begin{table}[ht]
    \centering
    \caption{Coefficients for Simulation Scenario M1: No Common Confounding}
        \begin{tabular}{|c|c|c|c|c| c|c| c|c|c|c|c|c|c|}
            \hline
             & Param. & $X_1$ & $X_2$ & $X_3$ & $X_4$ & $X_5$ & $X_6$ & $X_7$ & $X_8$ & $X_9$ & $X_{10}$ & $D_{1}$ & $D_{2}$\\
            \hline
            $D_1$ & $\boldsymbol{\beta}_{1}^{T}$  & 1 & 0.5 & 0.25 & 0.1 & 0.75 & 0 & 0 & 0 & 0 & 0 & - & -\\ 
            $D_2$ & $\boldsymbol{\beta}_{2}^{T}$   & 0 & 0 & 0 & 0 & 0 & 1 & 0.5 & 0.25 & 0.1 & 0.75 & - & - \\
            \hline \hline
            $Y$ & $\boldsymbol{\alpha}^{T}$     & 0 & 0.5 & 0 & 1 & 0 & 0.2 & 0 & 0 & 1 & 0 & 1 & 1 \\
            \hline
        \end{tabular}
        \caption*{All values of the covariates enter in each equation linearly with the respective coefficients shown in the table. In this scenario, covariates $X_{1}-X_{5}$ are associated with exposure $D_{1}$; however, only $X_{2}$ and $X_{4}$ are true confounders also associated with $Y$. Similarly, covariates $X_{6}-X_{10}$ are associated with exposure $D_{2}$; however, only  $X_{6}$ and $X_{9}$ are true confounders also associated with $Y$. There are no common confounders in this model. Each exposure has a true treatment effect coefficient of $1$.}
    \label{tab:m1_table}
\end{table}

\begin{table}[ht]
    \centering
    \caption{Coefficients for Simulation Scenario M2: Partially Common Confounding}
        \begin{tabular}{|c|c|c|c|c|c|c|c|c|c|c|c|c|c|}
        \hline
            & Param. & $X_1$ & $X_2$ & $X_3$ & $X_4$ & $X_5$ & $X_6$ & $X_7$ & $X_8$ & $X_9$ & $X_{10}$ & $D_{1}$ & $D_{2}$\\
            \hline
            $D_1$ & $\boldsymbol{\beta}_{1}^{T}$ & 0 & 0 & 1 & 0.5 & 0.25 & 0.1 &  0.75 & 0 & 0 & 0 & - & -\\ 
            $D_2$ & $\boldsymbol{\beta}_{2}^{T}$ & 0 & 0 & 0 & 0 & 1 &  0.5 & 0.25 & 0.1 & 0.75 &  0 & - & -\\
            \hline \hline
            $Y$ & $\boldsymbol{\alpha}^{T}$ & 0 & 0 & 0.5 & 0 & 0 & 1 & 0.2 & 0 & 1 & 0 & 1 & 1 \\
            \hline
        \end{tabular}
        \caption*{All values of the covariates enter in each equation linearly with the respective coefficients shown in the table. In this scenario, covariates $X_{3}-X_{7}$ are associated with exposure $D_{1}$; however, only $X_{3}$, $X_{6}$, and $X_{7}$ are true confounders also associated with $Y$. Similarly, covariates $X_{5}-X_{9}$ are associated with exposure $D_{2}$; however, only  $X_{6}$ $X_{7}$, and $X_{9}$ are true confounders also associated with $Y$. Common confounders of exposure $D_{1}$ and $D_{2}$ are $X_{6}$ and $X_{7}$. Confounder of exposure $D_{1}$ only is $X_{3}$, and $X_{9}$ is a confounder of $D_{2}$ only. Each exposure has a true treatment effect coefficient of $1$.}
    \label{tab:m2_table}
\end{table}

\begin{table}[ht]
    \centering
    \caption{Coefficients for Simulation Scenario M3: Common Confounding}
        \begin{tabular}{|c|c|c|c|c|c|c|c|c|c|c|c|c|c|}
        \hline
            & Param. & $X_1$ & $X_2$ & $X_3$ & $X_4$ & $X_5$ & $X_6$ & $X_7$ & $X_8$ & $X_9$ & $X_{10}$ & $D_{1}$ & $D_{2}$\\
            \hline
            $D_1$ & $\boldsymbol{\beta}_{1}^{T}$ & 1 & 0.5 & 0.25 & 0.1 & 0.75 & 0 & 0 & 0 & 0 & 0 & - & -\\ 
            $D_2$ & $\boldsymbol{\beta}_{2}^{T}$ & 0.8 & 0.8 & 0.05 & 0.4 & 0.55 & 0 & 0 & 0 & 0 & 0 & - & -\\
            \hline \hline
            $Y$ & $\boldsymbol{\alpha}^{T}$ & 0.5 & 0 & 1 & 0.2 & 1 & 0 & 0 & 0 & 0 & 0 & 1 & 1 \\
            \hline
        \end{tabular}
        \caption*{All values of the covariates enter in each equation linearly with the respective coefficients shown in the table. In this scenario, covariates $X_{1}-X_{5}$ are associated with exposure $D_{1}$ and $D_{2}$; however, only $X_{1}$, $X_{3}$, $X_{4}$ and $X_{5}$ are true confounders also associated with $Y$. Common confounders of exposure $D_{1}$ and $D_{2}$ are $X_{1}$, $X_{3}$, $X_{4}$, and $X_{5}$. There are no confounders of $D_{1}$ or $D_{2}$ only in this model. Each exposure has a true treatment effect coefficient of $1$.}
    \label{tab:m3_table}
\end{table}

\begin{table}[ht]
    \centering
    \caption{Covariate Balance}
    \begin{tabular}{|c|c|c|c|}
        \hline
        \textbf{Max Abs. Corr.} & \textbf{Avg. Abs. Corr.} & \textbf{ESS} & \textbf{Method}\\
\hline
 0.10 & 0.04 & 604 & mvGPS\\
\hline
0.19 & 0.09 & 578 & CBGPS (Macro)\\
\hline
0.21 & 0.08 & 672 & PS (Macro)\\
\hline
0.21 & 0.08 & 540 & Entropy (Macro)\\
\hline
0.35 & 0.11 & 768 & Entropy (Micro)\\
\hline
0.36 & 0.13 & 807 & PS (Micro)\\
\hline
0.41 & 0.15 & 685 & CBGPS (Micro)\\
\hline
0.41 & 0.20 & 1079 & Unweighted\\
\hline
    \end{tabular}
    \caption*{Assessing balance and effective sample size (ESS) using various weighted methods for the motivating example. Data used for estimating weights were restricted to $\mathcal{H}_{0.95}$. The univariate methods were applied separately to the two exposure metrics, macro or micro dose. The correlations are taken over both exposure metrics. The unweighted method represents the original values before applying weights. The weights for each method are trimmed using $q=0.99$.}
    \label{tab:cov_bal}
\end{table}
\clearpage

\clearpage

\clearpage

\clearpage

\clearpage

\clearpage

\clearpage

\clearpage

\clearpage

\end{document}